\documentclass[twocolumn,prc]{revtex4}
\usepackage{graphicx}
\begin{document}
\title{Interaction between Interpenetrating Charge Clouds and
  Collision of High-Energy Particles}
\author{A.Mukherji\footnote{Mailing address: D-111, Shanti Niket, 
 32/4 Sahitya Parishad Street, Kolkata, 700006, India} }
\affiliation{Formerly of \\ Saha Institute of Nuclear Physics, \\
        Bidhannagar, Kolkata, 700064, India }
\email{a.mukherji@saha.ac.in}
\begin{abstract}
Interaction between two interpenetrating spherically symmetric
charge distributions has been calculated. Limited range terms appear 
in addition to the Coulomb potential. Its strength increases and range
decreases with reducing sizes of the interacting particles. Between
two hydrogen atoms it yields the Morse Potential. Soft core potentials 
are obtained between pairs of nucleons. It has been shown that when
high-energy particles approach one another the potential between them
increases with increasing relative speed. They are not likely to 
disintegrate on impact. A possible way of smashing particles by
3-body collisions is indicated. 
\end{abstract}
\maketitle
\section{Introduction}
The equilibrium separation between the nuclei of two atoms in homopolar
bondage is usually less than the sum of their individual charge-radii.
The same is true for nucleons forming nuclei. In scattering experiments
the distance of closest approach of the particles gets even
smaller. Their charge volumes then interpenetrate or overlap. The
interaction between two spherically symmetric charge distributions in
such a configuration has been calculated. In addition to Coulomb 
potential short-range terms have been obtained. 

It has been observed that the strength and the range of the interaction 
are related to the sizes of the particles. From atoms to nucleons this
interaction grows million-fold in strength, while its range shrinks
from Angstrom to Fermi. 

The evidence relating to distributions of charge in protons and
neutrons have been evaluated. Two-component models with negative 
cores surrounded by positive clouds have been found suitable.
The potentials between various pairs of nucleons are nearly the same,
and this makes nuclear force appear charge-independent.

Particles approaching one another at high speed from opposite directions
undergo relativistic contractions. Their charge distributions get
highly condensed and the potential between them grows in strength.
The particles become harder and are unlikely to be decomposed by
direct collision. It is suggested that a particle at rest caught
between two speedy ones may be successfully smashed into its
elements.
\section{Overlapping Charges\label{overlapsec}}
\begin{figure}[]
\begin{center}
\includegraphics[scale=0.30]{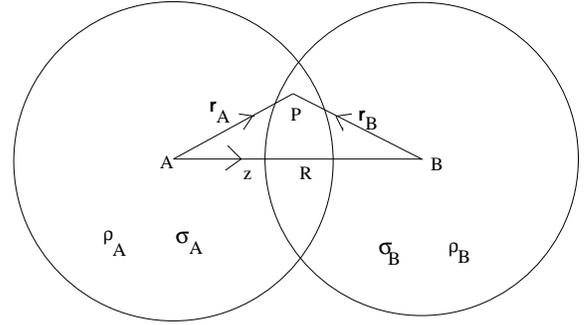}
\caption{Overlap of two extended charge distributions.
 \label{overlapfig}}
\end{center}
\end{figure}

Let two bodies $A$ and $B$ in Figure \ref{overlapfig} separated
by a distance $R$ carry spherically symmetric charge densities
$\rho_A(r_A)$ and $\rho_B(r_B)$ respectively. The potential\cite{jeans}
at a point $P$ due to $A$ is
\begin{equation}
\phi_P(r_A)~=~4\pi\left(\frac{1}{r_A}\int^{r_A}_0 \rho_A(r)r^2 dr +
\int^{\infty}_{r_A}\rho_A(r)rdr\right).  \label{overlapeq1}
\end{equation}

The interaction  energy between $A$ and $B$ is
\begin{equation}
V(R)~=~\int\phi_P(r_A)\rho_B(r_B)d\tau_B.   \label{overlapeq2}
\end{equation}
The features of $V$ depend upon the explicit forms of $\rho_A$
and $\rho_B$. 
\section{Overlapping Magnetic Moments  \label{overlapmagsec}}
Interaction between magnetic moments is weaker than that between
charges and may be ignored in the present context. But a brief 
sketch is relevant for the sake of completeness of the discussion
on overlap-dependent interactions. Some features of interest are 
mentioned below.

Let the bodies $A$ and $B$ of Figure \ref{overlapfig} carry
spherically symmetric magnetic moment densities $\vec{\sigma}_A(r_A)$
and $\vec{\sigma}_B(r_B)$ respectively. The magnetic field\cite{jeans}
at $P$ due to $A$ is
\begin{equation}
{\bf H}_P({\bf r}_A)~=~\frac{8\pi}{3}\vec{\sigma}_A(r_A) - 
 \frac{\vec{\mu}_A(r_A)}{r_A^3} + \frac{3{\bf r}_A}{r_A^5}
\vec{\mu}_A(r_A)\cdot {\bf r}_A  \label{magfldeq}
\end{equation}

The moment density $\vec{\sigma}_A(r)$ at any point $r$ within
a sphere of radius $r_A$ is considered to be the superposition
of a non-uniform density $\vec{\sigma}_A(r)-\vec{\sigma}_A(r_A)$
over a uniform background of strength $\vec{\sigma}_A(r_A)$,
the local density at radius $r_A$. $\vec{\mu}_A(r_A)$ represents
the contribution of the non-uniform part to the total moment
contained within the sphere of radius $r_A$.

The first term on the right hand side of Equation \ref{magfldeq} 
represents the field produced by the uniform background. 
The other two terms constitute the field due to the non-uniform 
component.

The interaction energy between $A$ and $B$ is
\begin{eqnarray}
V_M(R) & = & -\int \vec{\sigma}_B(r_B)\cdot {\bf H}_P(r_A)~d\tau_B 
  \nonumber \\
    & = & V_S + V_T,   \label{magenergy}
\end{eqnarray}
where $V_S$ and $V_T$ are its {\sl scalar} and {\sl tensor} components.

In an overlapping configuration $V_S$ is the dominant interaction.
It vanishes when the bodies are so far apart that they cease to
overlap and $V_T$ alone survives. Thus $V_S$ is the short-range
component and $V_T$ is the long-range one.

Between point particles $V_S$ reduces to Fermi expression \cite{fermi,
mukherji} involving a $\delta$-function and $V_T$ reduces to classical 
dipole-dipole interaction. It may also be noted that the tensor
component causes deuteron to have a quadrupole moment.
\section{H-H Potential \label{hhpot}}
For a pair of hydrogen atoms $A$ and $B$ (Figure \ref{overlapfig}) 
forming a molecule the electron charge densities may be obtained
from their ground state wavefunctions. Let them be given by
\begin{equation}
\rho_i(r_i)~=~C_i\exp(-\zeta_ir_i),~~i=A,B  \label{chargeform1}
\end{equation}
Their {\sl rms} radii are
\begin{equation}
\overline{r_i}~=~\sqrt{12}/\zeta_i
\end{equation}
and their charge contents are
\begin{equation}
q_i~=~8\pi C_i/\zeta_i^3.
\end{equation}

The interaction energy between $A$ and $B$ is
\begin{eqnarray}
V(R) & = & q_Aq_B\left[\frac{1}{R}-\frac{1}{R\left(\alpha^2-\beta^2
   \right)^3}\left[\beta^4\left(3\alpha^2-\beta^2\right)e^{-\alpha R}
  \right. \right.  \nonumber \\
  &   &  \hspace{3.5cm}\left.-\alpha^4\left(3\beta^2-\alpha^2\right)
  e^{-\beta R} \right]  \nonumber \\
  &  & \hspace{1.0cm} \left. -~\frac{\alpha\beta}{2\left(\alpha^2-
 \beta^2\right)^2} \left(\beta^3e^{-\alpha R} + \alpha^3e^{-\beta R}
 \right)\right] \label{hhpoteq1}
\end{eqnarray}
Here $\alpha$ and $\beta$ stand for $\zeta_A$ and $\zeta_B$ respectively.

In the limit $R~\rightarrow~0$, i.e. for total overlap
\begin{eqnarray}
V(0)  & =  &q_Aq_B~\frac{\alpha\beta}{2\left(\alpha+\beta\right)}
  \left[1+\frac{\alpha\beta}{\left(\alpha+\beta\right)^2}\right] \\
{\rm and}  &  &   \nonumber \\
V^{\prime}(0) & = & 0.
\end{eqnarray}
Hence $V$ is regular at $R=0$. Consequently, the self-energy of a
charged particle is {\sl not} infinite.

For $\alpha~=~\beta$
\begin{eqnarray}
V(R)~& = & ~q_Aq_B\left[\frac{1}{R} - \frac{e^{-\alpha R}}{R}\right.
  \nonumber \\
   &   & \left. -\frac{\alpha}{16}~e^{-\alpha R}\left(11+3\alpha R+
 \frac{1}{3}\alpha^2R^2 \right)\right].   \label{hhpoteq2} \\
{\rm and} &  &  \nonumber \\
V(0) & = & q_Aq_B\frac{5\sqrt{3}}{8~\overline{r}}  \label{v0eq}
\end{eqnarray}

The three terms on the right hand side of Equation~\ref{hhpoteq2} 
are the Coulomb potential and the {\sl classical equivalents} of Yukawa 
and exchange potentials respectively. The qualification implies that
Equation~\ref{hhpoteq2} has been obtained from classical considerations,
while the concept of exchange is fundamentally quantum mechanical.
A few comments on the situation may be pertinent at this stage.
When two charge clouds interpenetrate, the normal classical picture
is that they loose their individualities and form one composite
charge cloud. But in the present treatment it has been tacitly
assumed that the charge clouds could remember their past histories
and retain their identities while sharing the same space. Further,
a pure classical cloud, like a ball of dough, can split in
innumerable arbitrary fashions into a variety of components. But
when the combining clouds maintain their identities even in the
combined state, they split only in one way into their original
components. Their behaviour are no longer confined within the
realm of pure classical physics. The situation is somewhat
similar to solitons passing through one another. They do not form
a single entity when they get mixed up. Hence results beyond the
domain of technically classical physics may not be unexpected
in the preent context.

The Coulomb potential in Equation~\ref{hhpoteq2} is long-ranged and the
rest constitute a short-ranged interaction. The latter dominates
over the former so long as there is significant overlap of the charge 
clouds of the particles. Hence, its range is of the
order of the charge radii of the interacting bodies.

If $B$ is so small compared to $A$ that it may be treated as a point
particle,
\begin{equation}
V(R)~=~q_Aq_B \left[\frac{1}{R} -\frac{e^{-\alpha R}}{R}-\frac{\alpha}{2}
  e^{-\alpha R}\right].  \label{hhpoteq3}
\end{equation}
It reduces to pure Coulomb potential when $A$ and $B$ are widely
separated and the particles can be treated as points.

For a pair of hydrogen atoms Equation~\ref{hhpoteq2} may be used for the
interaction between the electron clouds of the atoms. Equation~
\ref{hhpoteq3} gives the interaction of the nucleus of one atom with 
the electron cloud of the other atom. The nuclei themselves interact
through the Coulomb potential. The resulting potential agrees with the 
Morse potential\cite{slater}
\section{Nuclear Potential \label{pppot}}

It may be noted that $V$ is stronger for smaller particles. If the
{\sl rms} radii get reduced, $V(0)$ (Equation \ref{v0eq}) gets
proportionately enhanced. Since the radius of a nucleon is smaller
than that of an atom by a factor of $10^5$, the potential, which is
of the order of {\sl eV} between atoms, becomes of the {\sl MeV} order 
between nucleons\cite{mukherji2}.

Two protons $A$ and $B$ (Figure \ref{overlapfig}) in close proximity
have their quark charge clouds interpenetrating one another. The charge
densities may be expressed as
\begin{equation}
\rho_i(r_i)~=~C_0\exp(-\alpha r_i)~+~C_1\beta r_i\exp(-\beta r_i),
  ~~~i=A,B   \label{ppchargeeq}
\end{equation}
The {\sl rms} radii are
\begin{equation}
\overline{r_0}~=~\sqrt{12}/\alpha,~~~{\rm and}~~~\overline{r_1}~=~
  \sqrt{20}/\beta.
\end{equation}
The corresponding charge components are
\begin{equation}
q_0~=~8\pi C_0/\alpha^3,~~~{\rm and}~~~q_1~=~24\pi C_1/\beta^3.
\end{equation}
They are subject to the restrictive condition that the total charge of
proton  
\begin{equation}
e_P~=~q_0+q_1.
\end{equation}

\begin{figure}[]
\begin{center}
\includegraphics[scale=0.70]{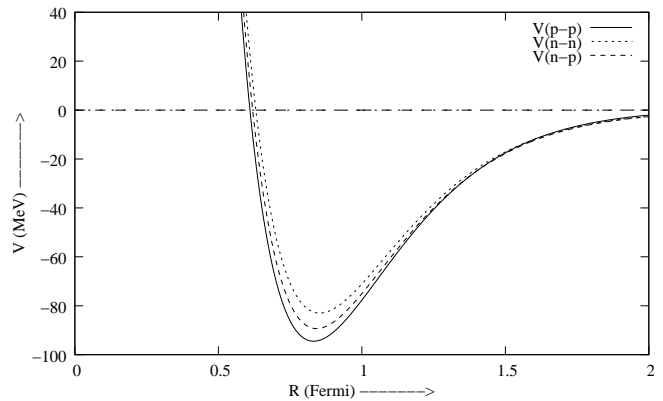}
\caption{Plot of internucleonic potentials V in units of {\sl MeV}
as a function of the internucleonic separation R in units of {\sl fm}
with $\bar{r_0}$ = 0.005 \& $\bar{r_1}$ = 0.8.  \label{potsfig}}
\end{center}
\end{figure}

Using the values $\overline{r_0}=0.005,~~\overline{r_1}=0.8$ in units of
{\sl fm} and $q_0=-32,~~q_1=+33$ in units of $e_P$, the calculated
p-p potential is shown in Figure \ref{potsfig} as $V(p-p)$. It is in fair
agreement with the soft core potential obtained by Reid\cite{reid} 
from scattering data.

The charge distribution in proton is conventionally taken to be 
positive throughout its volume\cite{littauer}. This is incompatible 
with a potential which is strongly repulsive in one region and
attractive in another. But that does not create any conceptual
difficulty since potential and charge are considered to be 
dissociated from each other. The force between nucleons is known
not to depend upon the charges carried by them. The {\sl natural
conclusion} is that interaction between charges is
not related to the strong force. But still one link remains.
Quark hamiltonian involves the potential and quark wavefunctions
produce the charge densities. The two cannot, therefore, be
totally alienated. There lies an inconsistency. The question is
whether the {\sl natural conclusion} is the only possible conclusion.
Before searching for an answer to this question, a closer view of
the all-positive distribution of charge in proton is necessary.

A proton with +1 unit of charge in the nucleus of a heavy atom can 
capture a K-shell electron
carrying -1 unit of charge. The additional -1 unit does not 
nullify the charge distribution in the proton. Neither should it
modify the existing distribution to any great extent. It should primarily 
affect the distribution of charge in the peripheral region.
Otherwise, the reverse process, i.e. the spontaneous decay of a
free neutron with a definite half-life, will be hard to conceive.
An all-positive proton requires half of its total charge to change
sign in order to turn itself into a neutron. This is provided by
the charge of the captured electron. The -1/2 unit
of charge has to reverse itself into +1/2 unit in order to turn
the neutron back into proton. It is enabled by emission of an
electron. Eithe way the process involves a 200\% change for 50\%
of the total charge of proton. Such a  drastic reshuffling of the 
distribution of charge associated with the
decay of a neutron under no external influence or transfer of energy is
rather unrealistic. A proton will be able to accommodate an electron
only if its addition does not seriously disturb the original
charge distribution of the proton. The electron charge should, therefore,
remain confined beyond the charge volume of a proton.A neutron should, 
therefore, appear to have one positively charged sphere of one unit
embedded in a negatively charged shell also of one unit. Such a 
structure has not been substantiated independently.

Since n-p and p-p potentials are nearly the same, the quark
wavefunctions in neutron and proton may be expected to be similar
in nature. Two greatly different charge distributions in them
are inconsistent.

An {\sl ab initio} calculation of charge densities in proton and
neutron requires determination of quark wavefunctions in them. In
absence of a dominating central potential, it calls for
self-consistent solutions of Schr\"{o}dinger equations. The
project has not been undertaken at present.

Another problem is that the all-positive charge distribution in
proton fails to account for {\sl back-scattering} of electrons.
The presence of a negative core in proton would have eased the
problem.

An alternate outlook may be considered. It may be accepted
that an interaction between overlapping charges produces
strong force as per the presentation made hereinbefore.
A negative core may be added to a magnified positive outer
distribution in a proton to produce a net charge of +1 unit. The two
charge components may individually be much larger than 1 unit,
so that the gain or loss of one -1 unit does not affect
the overall charge distributions severely. One can then
expect the {\sl rms} charge radii of proton and neutron to
be not much different. Nuclear force may then appear to be
{\sl charge independent}. The present model could provide an
alternative to the {\sl natural conclusion} mentioned before.
Such interplay of large quantities
of charges is not unknown. All atoms are electrically
neutral when observed from a distance, while closer
scrutiny of heavy atoms reveals large amounts of positive
and negative charges in co-existence.

The charge distribution in a proton as considered here has a negative
core of 32 units surrounded by an outer positive cloud of
33 units. 
\begin{figure}[]
\begin{center}
\includegraphics[scale=0.70]{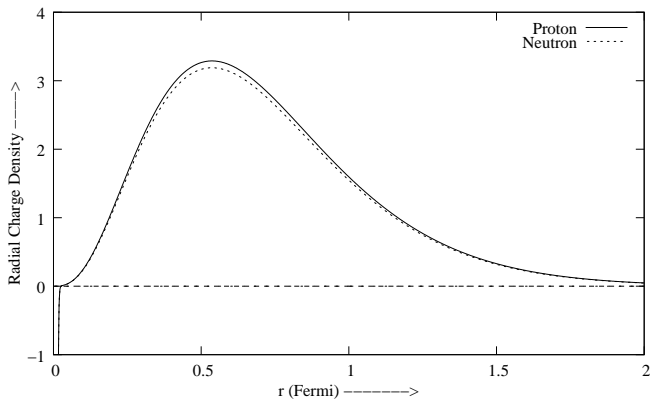}
\caption{Plot of $r^2\rho(r)$ for proton and neutron in
arbitrary units as a function of r in units of {\sl fm}
with $\bar{r_0}$ = 0.005 \& $\bar{r_1}$ = 0.8.  \label{chargefig}}
\end{center}
\end{figure}
The neutron core carries the same charge as the proton core, but
its outer cloud has only +32 units of charge.
The radial charge densities are shown in Figure~\ref{chargefig}.
The difference is only 1 unit
out of 32 units in the outer charge cloud. It amounts  to a
3\% change, which is well within acceptable limits. Such a small
change does not put up an objection to a qualitative statement.
Various internucleonic potentials are plotted in Figure~\ref{potsfig},
which again bear close resemblance. They confirm the expectations
mentioned before that the {\sl rms} charge radii of proton and
neutron are nearly equal and that nuclear force is {\sl charge
independent}.

It needs be mentioned that the numbers cited above are not
sacrosanct. They were chosen in order to demonstrate that a
suitable quark charge density distribution can produce a
potential between two protons in agreement with experimental
findings. Quark wavefunctions obtained {\sl ab initio} may
confirm or modify the present choice of the parameters. Two
terms in Equation~\ref{ppchargeeq} is the minimum requirement
and is adequate for the present purpose. A broader basis set 
may offer better representation of quark charge densities in
nucleons. But the basic contention of the present analysis
remains unchanged.

The negative core may also be investigated by positron
scattering from protons. Such experiments would require
positron beams of sufficient intensity with adequate energy
to ride over or tunnel through a potential hill before
they reach the core.
\section{High-energy Collisions}
\begin{figure}[]
\begin{center}
\includegraphics[scale=0.30]{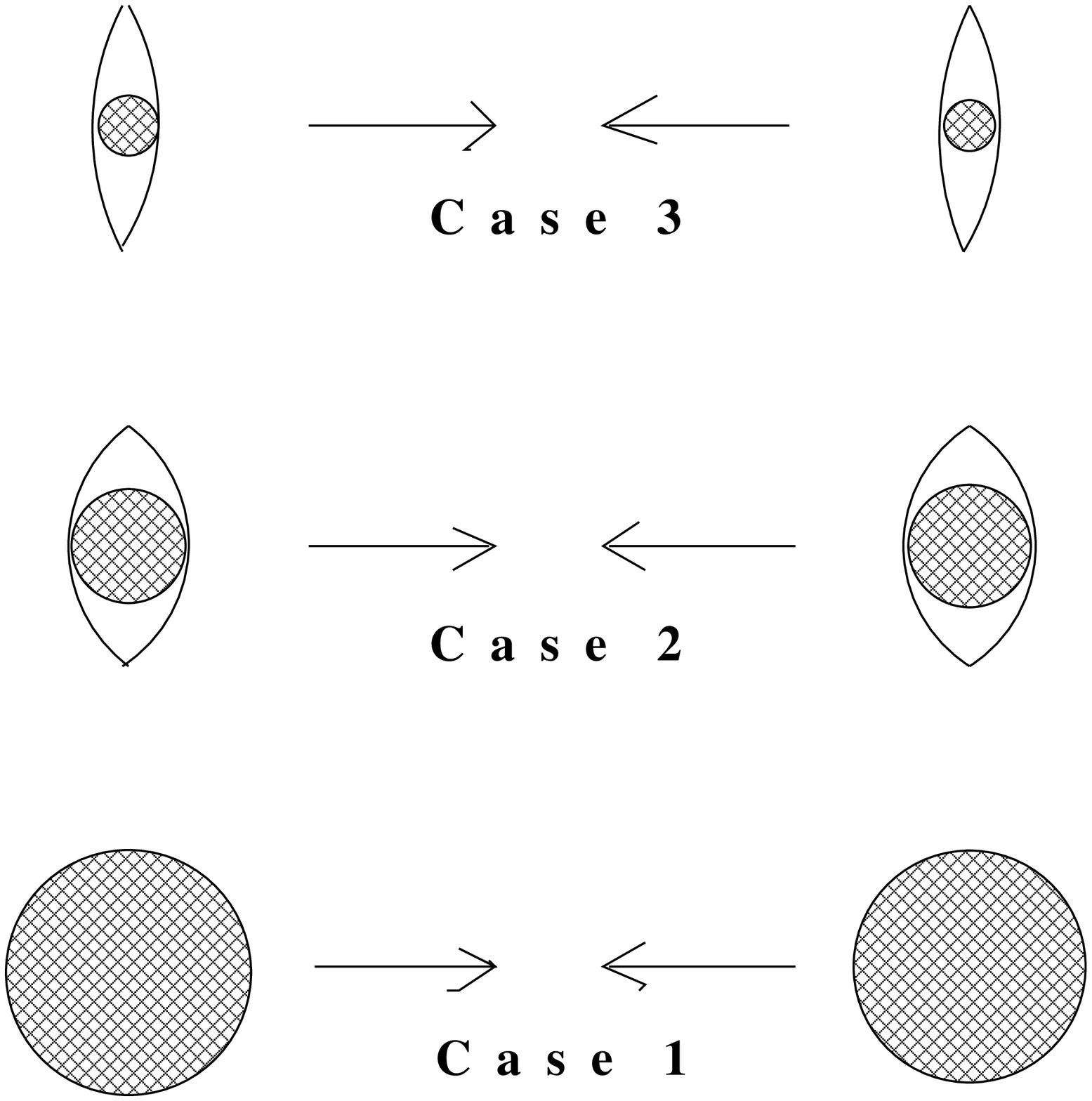}
\caption{Schematic plot of two colliding particles. Case 1 corresponds
to non-relativistic relative velocities. Case 2 corresponds to
moderate relative velocities: particles suffering 50\% contractions.
Case 3 corresponds to high relative velocities: particles suffering
75\% contractions. The balls are indicated by the hatched inscribed
circles. The rest of the material constitute the rings.
  \label{collfig}}
\end{center}
\end{figure}

Observed from the centre-of-mass, two protons colliding with
relativistic energies appear to undergo contractions along
their line of approach. They get flattened out(Figure~\ref{collfig}).
The shape of such a proton may be approximated by a ring with
a ball just fitting into its hole. Its mass and charge are now shared
between the ring and the ball.

With increasing energy, it seems that the proton is further
squeezed. The ring gets flatter and wider so as to reduce the
size of the hole while preserving its outer diameter. The ball
now closing the hole is smaller. The total volume of the
proton gets reduced. Naturally the mass and the charge 
distributions get denser and their partitioning between the
two components vary. However, since the charge density in a
proton tails off rapidly, the ring  carries very little charge.

\begin{figure}[]
\begin{center}
\includegraphics[scale=0.65]{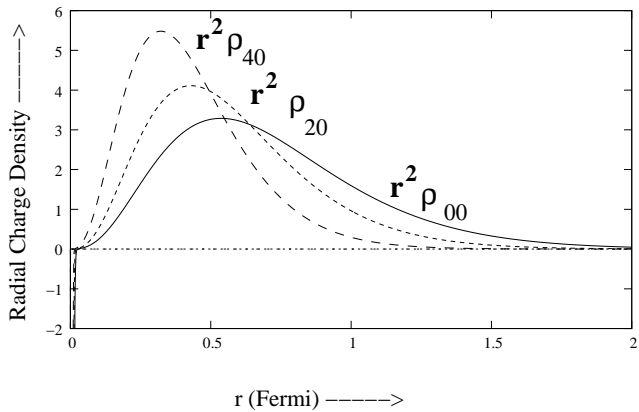}
\caption{Plot of $r^2\rho(r)$ in arbitrary units as a function of r 
in units of {\sl fm}.  For $r^2\rho_{00}$,~~ $\bar{r_0}$ = 0.005 \& 
$\bar{r_1}$ = 0.8. For $r^2\rho_{20}$ these have the values 0.004 
\& 0.64 respectively and for $r^2\rho_{40}$ these values are 
0.003 \& 0.48.  \label{r2rhofig}}
\end{center}
\end{figure}

\begin{figure}[]
\begin{center}
\includegraphics[scale=0.70]{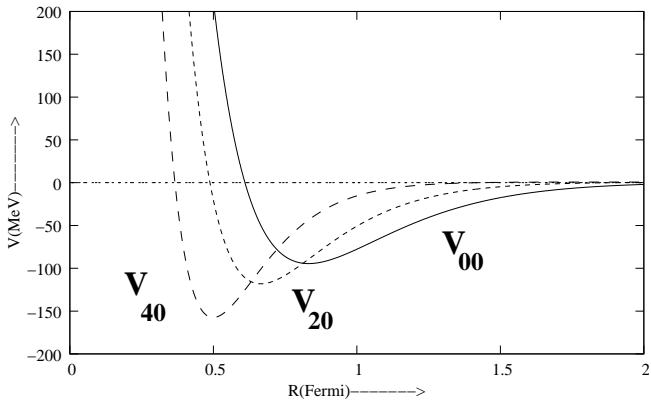}
\caption{Plot of the interparticle potential V in units of {\sl MeV}
as a function of the interparticle separation R in units of {\sl fm}.
For $V_{00}$ $\bar{r_0}$ = 0.005 \& $\bar{r_1}$ = 0.8. For
$V_{20}$ these have the values 0.004 \& 0.64 respectively and for
$V_{40}$ these values are 0.003 \& 0.48. \label{potfig}}
\end{center}
\end{figure}

In the following qualitative analysis the ring is ignored. The modified
scenerio is that with increasing energy a spherical proton
remains spherical but gets smaller. The densities of mass and 
charge increase. The proton thus gets more and more compact.
$r^2\rho_{20}$ and $r^2\rho_{40}$ in Figure~\ref{r2rhofig} show
the radial charge densities at $20\%$ and $40\%$ reductions of
the {\sl rms} radii respectively. Compared to the unreduced
density $r^2\rho_{00}$ they are laterally contracted and
vertically elongated.
 Corresponding $p-p$ interaction potentials 
$V_{20}$ and $V_{40}$ are plotted along with $V_{00}$, the
non-relativistic potential, in Figure~\ref{potfig}.
 They also display similar
deformations. The process continues with decreasing size of the
proton associated with increasing energy.
The peak of the potential is
attained at $R=0$. It is about $4\times 10^5MeV$ for a proton
at rest or with non-relativistic energy.

Since the interaction between two reducing spheres goes on
changing, the data collected from scattering experiments
performed at different relativistic energies cannot be
correlated unless the change of potential with energy is
taken into consideration. 

With increasing energy of collision the participants become 
more massive, very compact and extremely hard. Their 
fundamental ingredients are glued together more strongly. 
They are reluctant to leave their progressively reducing
confines. Consequently, the colliding particles strongly
resist disintegration under direct impact. Relativistic 
effects are comparatively less pronounced for a heavy
nucleus carrying higher momentum with lower velocity.
Collision between them may result in splintering into
their constituent particles rather than totally
disintegrating into their basic elements. 

A 3-body collision may be induced by injecting a proton
plasma or hydrogen atoms between two beams of colliding
nuclei. From the centre-of-mass such a proton appears
stationary. It is {\sl `soft'}. The high speed nuclei are hard
hitters. Some of the soft protons may get caught between
them and be smashed.

If proton beams are used, the probability of a head-on
collision gets reduced. A nucleus with mass number $M$
has a cross-section $M^{2/3}$ times that of a proton.
The probability of two such nuclei colliding is $M^{4/3}$
times that of two protons.

The soft protons injected into the collision chamber
must neither be too dense nor too thin. In a thick cloud
most of the beam protons will hit plasma protons in 2-body
collisions. From the centre-of-mass both will appear to 
be moving at half the beam proton's speed. It will result
in a collision between two hard protons. On the other hand,
if the plasma is too thin, the colliding protons will find it
difficult to get a soft proton in-between them.

Slower protons are preferable. They spend more time in the 
collision chamber and have better opportunities of
capturing soft protons. But they must be energetic
enough to smash the captive ones. Even then, 3-body
events may occur infrequently and a successful experiment
will require sensitive detector systems. 

The ultimate result of a collision experiment depends upon the 
answer to yet another question that still
persists: {\sl `what happens when a soft proton is smashed?'} 
Do the elementary particles constituting it fly off
in different directions, or does it get flattened out
and subsequently bounces back into its original form?
Successful decomposition of a proton depends upon how much concussion 
the gluonic binding can withstand.
\section*{Acknowledgement}
The author is indebted to Prof. P.Rudra, formerly of University of
Kalyani, for very valuable assistance.

\end{document}